%% file: main.tex
\documentclass[a4paper,11pt]{article}
\pdfoutput=1
\usepackage{jinstpub}
\usepackage{lineno}
\usepackage[separate-uncertainty]{siunitx}
\usepackage{subcaption}
\usepackage{adjustbox}

\title{Evaluation of polymer-metal-hybrid bonded wafer-stacks and sensor wafers for ultra-thin hybrid silicon detectors}

\author[a,1]{Janna Zoe Vischer,\note{Corresponding author}}
\author[b]{Yannick Dieter, }
\author[b]{Jochen Dingfelder, }
\author[c]{Thomas Fritzsch, }
\author[b]{Fabian Hügging, }
\author[a]{Kevin Kröninger, }
\author[b]{Maximilian Mucha, }
\author[b]{Matthias Schüssler }
\author[a]{and Jens Weingarten}

\affiliation[a]{Department of Physics, TU Dortmund University, 44227 Dortmund, Germany}
\affiliation[b]{Physikalisches Institut, Rheinische Friedrich-Wilhelms-Universität Bonn, 53115 Bonn, Germany}
\affiliation[c]{Fraunhofer IZM, 13355 Berlin, Germany}

\emailAdd{janna.vischer@tu-dortmund.de}

\input{abstract.tex}

\begin{document}
\maketitle
\flushbottom

\section{Introduction}
\label{sec:intro}
\input{introduction2.tex}

\section{Hybrid Wafer-to-Wafer bonding of Daisy-Chain Wafers}
    \label{sect:W2WbondingDCW}
    \subsection{Production and Properties of the Daisy-Chain Wafer stack}
        \label{subsect:DCW}
        \input{DCW.tex}
    \subsection{Bond Quality Measurements}
        \label{subsect:BondQuali}
        \input{BondQuali.tex}

\section{Characterization of sensor wafers prior to hybridization}
    \label{sect:SW}
    \input{SensorWafer.tex}
    
\section{Conclusion and perspective}
\label{sect:Conclusion}
\input{Conclusion.tex}

\section*{Acknowledgements}
The authors acknowledge the usage of the wafer prober at the TU Dortmund. Funded by the Deutsche Forschungsgemeinschaft (DFG, German Research Foundation)-450639102. 
This work was supported by the BMBF-Verbundprojekt „05H2021 - R\&D DETEKTOREN (Neue Trackingtechnologien): Entwicklung aktiver und passiver mikrostrukturierter CMOS-Sensoren“ (Grant number Bonn University: 05H21PDR1, TU Dortmund University: 05H21PERD1).

\bibliographystyle{JHEP}
\bibliography{biblio.bib}

\end{document}

%% file: abstract.tex
\abstract{
Semiconductor pixel detectors are widely established in High Energy Physics (HEP) and Medical physics for their high spatial resolution and tracking capabilities. Research on both monolithic detectors and hybrid detectors is ongoing. 
Monolithic detectors, which integrate the sensor and the read-out electronics in the same die, provide the benefit of reduced thickness but the needed intricate imaging process is only offered by a limited number of chip vendors.
The hybrid approach instead facilitates the design and fabrication of sensor and read-out chip using different technologies and opens up access to a large market of semiconductor vendors. 
\\
For the production of silicon pixel detectors, the interconnection between sensor and read-out chip is usually realized on an individual die level. The needed mechanical stability during the handling of the dies limits their possible thinness.
The wafer-to-wafer interconnection process being developed in this project uses a polymer underfill layer between the wafers to provide additional mechanical stability.
This allows one to thin the wafer stack significantly after interconnection, bringing the total thickness close to that of monolithic detectors.
In this paper, we present first results on the bump bonding yield of the process based on daisy-chain wafer measurements.\\
For the first hybrid pixel detectors produced with this technique, a dedicated sensor wafer was designed and fabricated to be bonded to Timepix3 read-out chip wafers. 
Results of the characterization of the sensor wafer before hybridization are presented.
We show that the wafer-to-wafer bonding process is suitable for hybrid semiconductor pixel detectors.
}

\keywords{Instrumentation for high-energy physics, hybrid detectors, hybridization}

%% file: introduction2.tex
Most current tracking detectors in high-energy physics use silicon pixel detectors in their innermost layers.
Their excellent spatial resolution and radiation tolerance make them very suitable for the typical harsh environments close to the interaction points of both collider and fixed-target experiments~\cite{Wermes_2015}.
Research and development is ongoing to further improve both hybrid detector concepts as well as monolithic active pixel sensors (MAPS).
MAPS promise excellent spatial and time resolution in certain regimes, as well as the potential for large-scale production of detectors at commercial CMOS foundries~\cite{Snoeys_2023}.
However, the number of semiconductor vendors offering suitable CMOS processes for radiation-hard particle detectors is limited.
Hybrid semiconductor pixel detectors consist of separate sensors and read-out chips that need to be bonded together. This enables flexibility in the development of both parts and reusability of designs in different settings. 
However, hybridization is currently done on die level, introducing significant processing time and cost.
In addition, the traditional handling of individual sensor and read-out dies requires a certain die thickness for stability, which increases the overall radiation length of the detector.
Therefore, advanced interconnection techniques for hybrid detectors are of increasing interest. One such technique is wafer-to-wafer bonding in which the sensor and read-out chip are connected on the wafer-level. This interconnection techniques preserve the flexibility to design sensor and read-out chip separately, while enabling large-area fabrication and testing at wafer level. An additional advantage for applications in high-energy physics is the possibility to thin the wafer stack to a suitable level in terms of the total radiation length.
\\
\\
\noindent
Wafer-to-wafer bonding technologies are already widely used in the mass production of advanced 3D stacked devices, such as mobile phone cameras~\cite{NovelStackedCMOS}.
There, the electrical interconnects between the two wafers are fabricated using a metal-oxide hybrid bonding process. 
However, in addition to IP licensing fees, this technology requires state-of-the-art process tools and wafer surfaces with a planarity of a few nanometers.
A polymer-metal-hybrid bonding technology has been developed by Fraunhofer IZM to overcome these special requirements.
It must provide reliable electrical pixel-to-pixel interconnections and mechanical stability for handling the ultra-thin wafers.
Additionally, sufficient stability must be achieved during high-temperature and vacuum processing steps, particularly during silicon etching for through-silicon via (TSV) interconnects or wafer-to-wafer bonding.
\\
\\
\noindent
In this paper, we present the results of first fundamental tests towards a fully wafer-to-wafer bonded silicon detector.
In Section~\ref{sect:W2WbondingDCW}, we present wafer-to-wafer interconnection process and quantify the interconnection yield using two daisy-chain wafers (DCW) to determine the suitability of the technique for the production of hybrid detectors.
Section~\ref{sect:SW} introduces the next step of the project.
A sensor wafer has been designed and produced to be bonded with a Timepix3 read-out wafer. 
Two sensor wafers are characterized in terms of depletion and breakdown voltages to verify their suitability for realizing ultra‑thin, low‑mass hybrid pixel detectors~\cite{AIDAinnova}. The paper concludes in Section 4 where the next steps towards a fully functional detector is outlined.

%% file: DCW.tex
\begin{figure}
    \centering
    \includegraphics[width=0.4\linewidth]{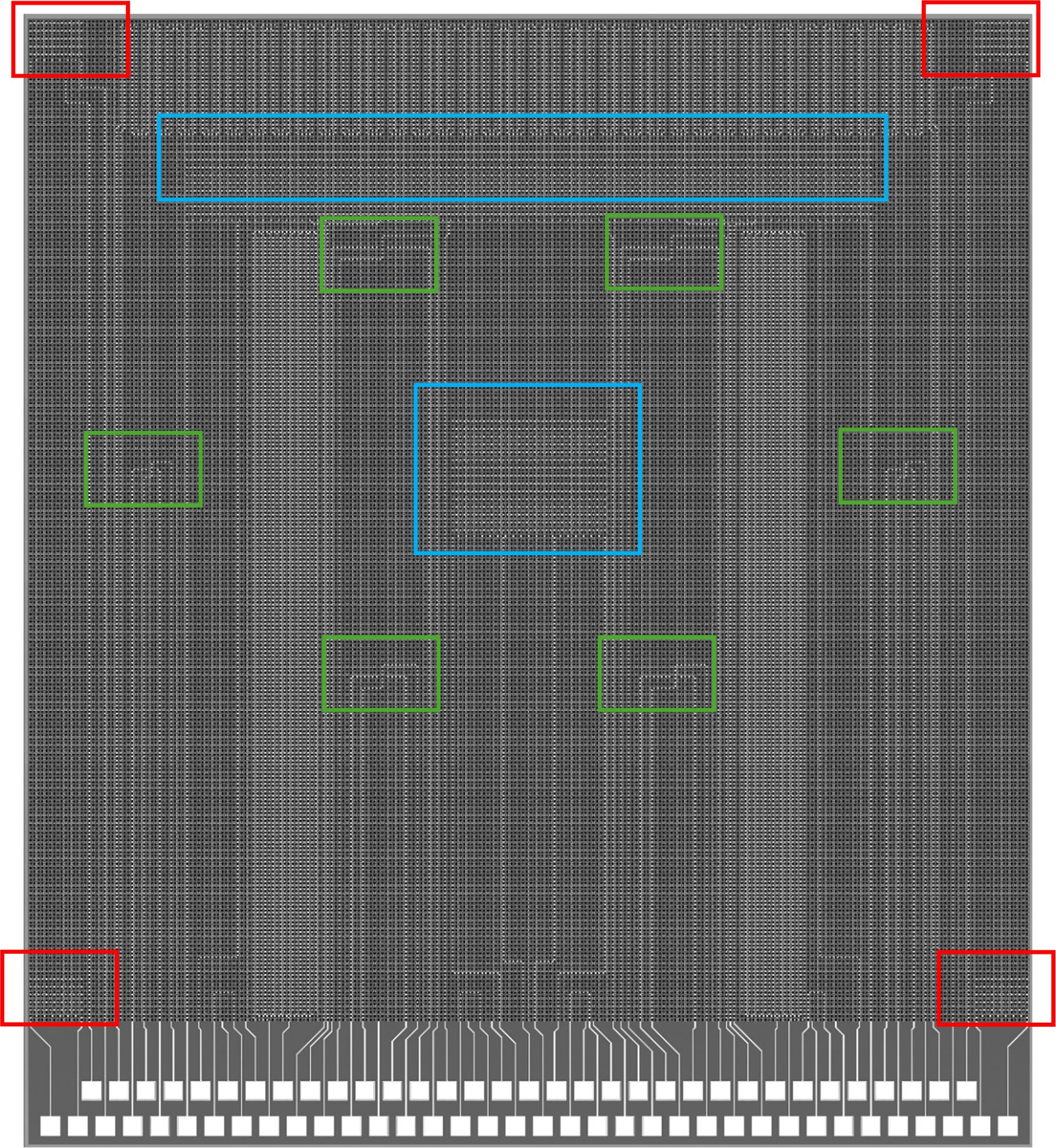}
    \caption{Position of the different test structures on a DCW die. In green six single bond structures are shown. In red the corner daisy chains are highlighted. The long daisy chains in the center and top of the die are shown in blue. The probe pads are located at the lower end of the die.}
    \label{fig:Teststructures}
\end{figure}
\begin{figure}
    \begin{subfigure}[]{0.45\textwidth}
        \centering
        \includegraphics[height=0.8\linewidth]{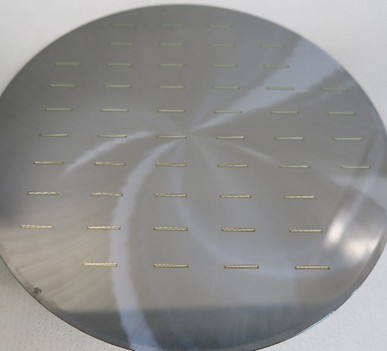}
        \caption{ }
        \label{fig:DC_Wafer_full}
    \end{subfigure}
    \begin{subfigure}[]{0.45\textwidth}
        \centering
        \sbox0{\includegraphics[height=0.5\linewidth, clip=true]{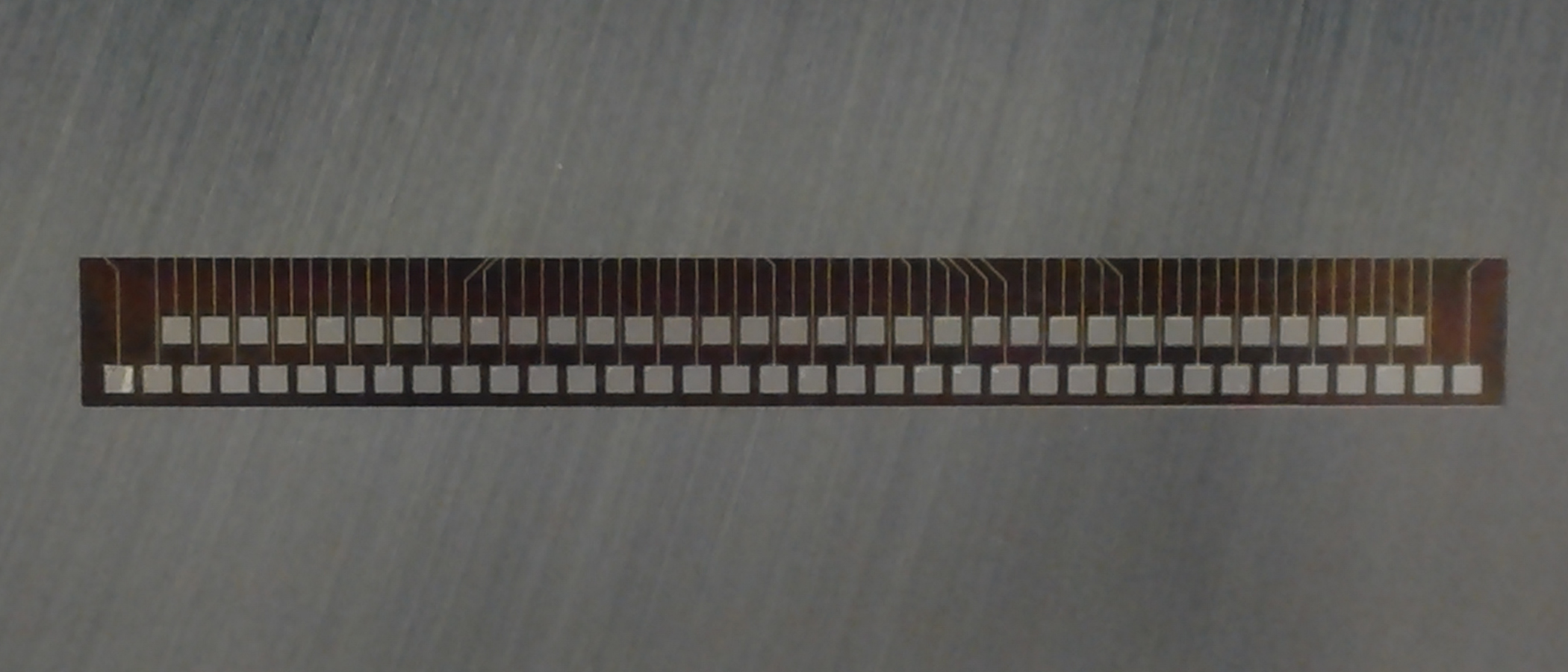}}
        \includegraphics[height=\ht0,keepaspectratio]{Contacs.png}
        \caption{}
        \label{fig:etched_window_to_pads}
    \end{subfigure}
    \caption{Photos of the DCW stack. \autoref{fig:DC_Wafer_full} shows the full \SI{200}{\milli \meter} wafer stack with its \num{50} etched windows to access the probe pads with probe needles located on the lower wafer. \autoref{fig:etched_window_to_pads} shows a close-up of such an etched window in the top wafer of the wafer stack with the square probe pads.}
    \label{fig:Wafer_photos}
\end{figure}
\noindent 
The daisy-chain wafer is a dedicated \SI{200}{\milli \meter} process‑development wafer, designed with individual probe test structures and daisy chains. 
It serves as the platform for the systematic evaluation of the performance after wafer-to-wafer bonding.
The daisy-chain chip design with the individual test features is shown in \autoref{fig:Teststructures}.
This test chip architecture allows for direct resistance measurements of individual interconnects through single bond structures shown in green. 
Additionally, four corner daisy chains are shown in red and two long daisy chains in the center and top of the chip in blue. 
They enable the qualitative and quantitative assessment of the bonding yield and uniformity across the bonded DCW stack.
\\
\begin{figure}
\centering
\begin{subfigure} [c]{0.45\textwidth}
    \centering
    \includegraphics[width=0.9\linewidth]{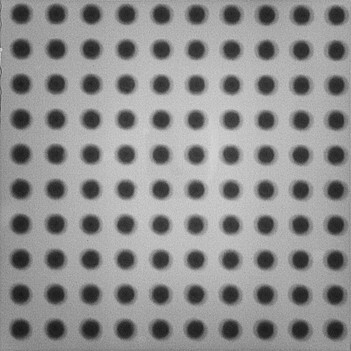}
    \caption{}
    \label{fig:DCW_xray}
\end{subfigure}
\begin{subfigure} [c]{0.45\textwidth}
    \centering
    \includegraphics[width=\linewidth]{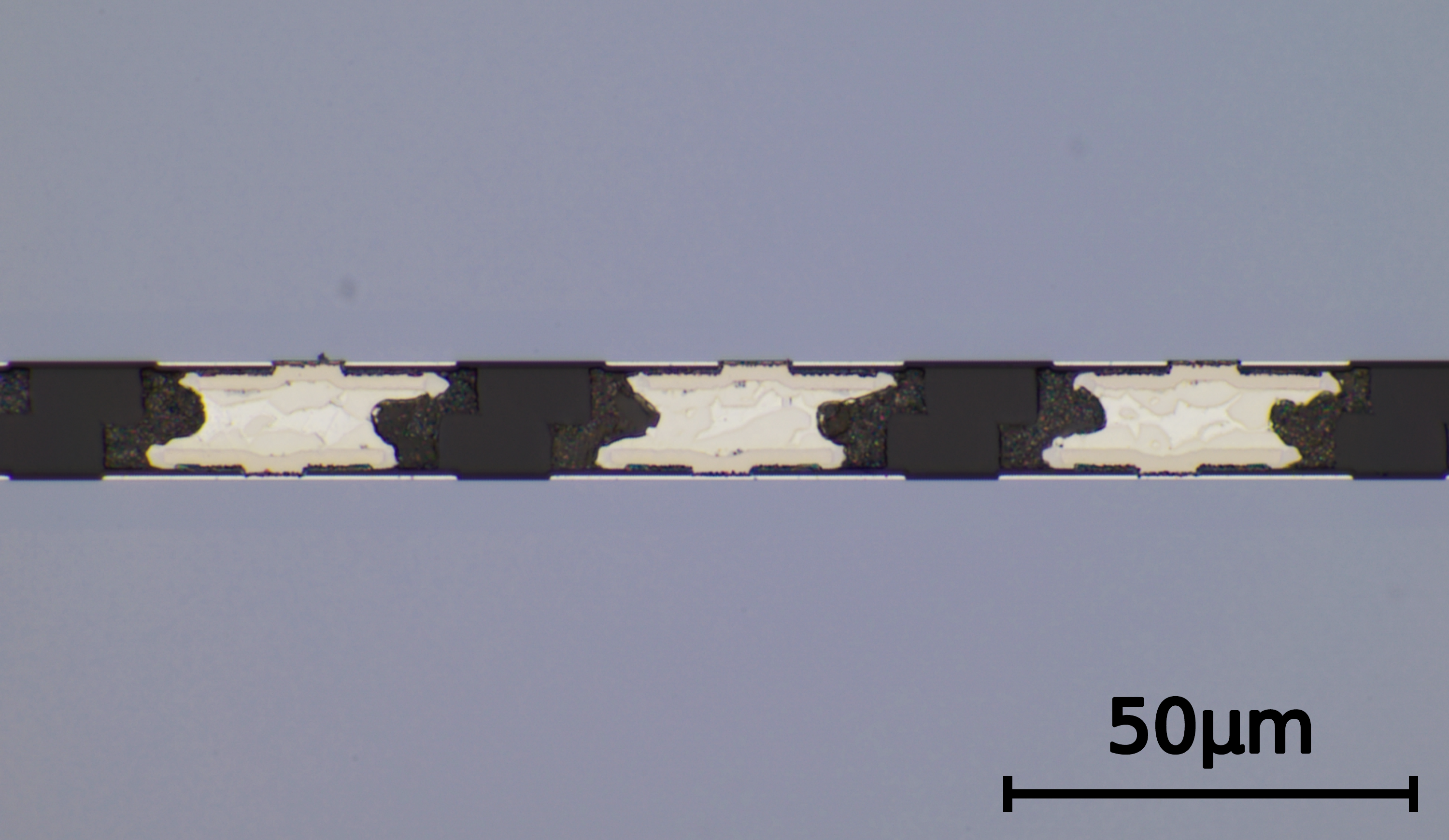}
    \caption{}
    \label{fig:DCW_crossection}
\end{subfigure}
    \caption{X-ray image and cross section of DCW stack after wafer-to-wafer bonding. The X-ray image in \autoref{fig:DCW_xray} shows well aligned solder pillars.
    \autoref{fig:DCW_crossection} shows a cross section of a DCW stack with good electrical interconnection (golden) between top and bottom wafer (blue gray). A slight misalignment of approx. \SI{4}{\micro \meter} is visible.}
    \label{fig:DCW_alignment}
\end{figure}
\\
\noindent
After the fabrication of the process-development wafer and the selection and evaluation of several polymeric wafer bond materials, a polymer-metal-hybrid bonding process has been developed by Fraunhofer IZM.
Key features of the evaluated process are the deposition of Ni-Cu-SnAg solder pillars, the deposition and patterning of the polymeric bond layer on top and bottom wafer, followed by chemical mechanical polishing to achieve a highly uniform, planar bonding interface.
Subsequent wafer-to-wafer bonding achieves a robust metal-polymer hybrid bond. After bonding, the top wafer is thinned to \SI{50}{\micro \meter} and openings are etched to access probe pads, enabling electrical characterization of the bonded structures.
\autoref{fig:DC_Wafer_full} shows the finished DCW stack and \autoref{fig:etched_window_to_pads} a close-up of the etched window and the probe pads.
\\
\\
\noindent
X‑ray microscopy demonstrates excellent wafer alignment, with center‑to‑center misalignment below about \SI{4}{\micro \meter} shown in \autoref{fig:DCW_xray}.
Cross‑section analysis confirms continuous intermetallic connections between the top and the bottom wafer as shown in \autoref{fig:DCW_crossection}. 

%% file: BondQuali.tex
The DCW stack was investigated in a semi-automatic wafer prober MPI TS3500-SE. 
Its controlled environment ensures a stable temperature slightly above room temperature between \SI{25}{\celsius} and \SI{26}{\celsius}. 
Its housing provides shielding from ambient light and reduces electromagnetic noise. 
The Python API of the wafer prober combined with the remote control of the used multimeter and power source allows for automated measurements of the whole wafer including stepping from die to die. 
Before the measurements, the topography of the wafer was optically determined using the wafer prober station to ensure proper contact of the probes. 
The probe needles were positioned manually for each measurement series. 
In between measurement series, the probe needles were cleaned of collected debris in regular intervals using a swab dipped in isopropanol.
\\
\\
\noindent
The resistances of the structures on the DCW stack were measured using a Keithley~2000 multimeter. 
For each die, the resistance was measured ten times without moving the needles before automatically stepping to the next die. 
The statistical and systematic uncertainties depending on the measurement range of the multimeter were taken into account and the mean resistance of each bond was calculated.
All measurements were performed with a four-point measurement setup to exclude the resistance of the cables from the multimeter to the probe needles.

\subsubsection{Single Bonds}
\label{subsubsect:SingleBonds}
\begin{figure}[htbp]
    \centering
    \begin{minipage}{0.3\textwidth}
        \begin{subfigure}{\linewidth}
            \centering
            \includegraphics[width=\linewidth]{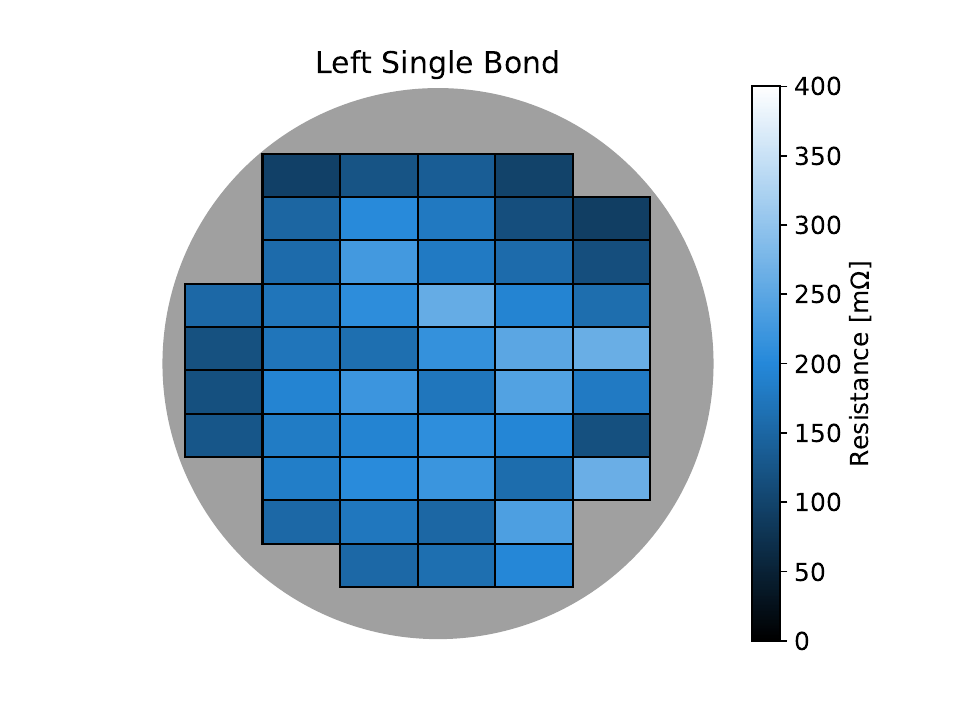}
            \caption{}
            \label{subfig:SBSLeftCommScale}
        \end{subfigure}
    \end{minipage}
    \hfill
    \begin{minipage}{0.3\textwidth}
        \begin{subfigure}[]{\linewidth}
            \centering
            \includegraphics[width=\linewidth]{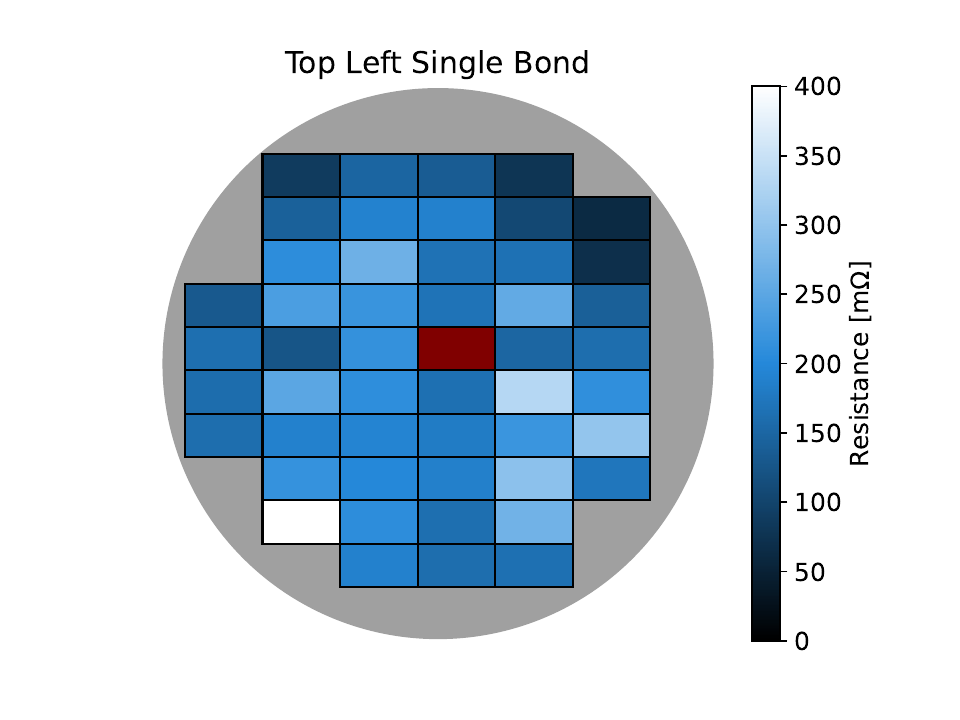}
            \caption{}
            \label{subfig:SBSTopLeftCommScale}
        \end{subfigure}
        \bigskip
        \begin{subfigure}[]{\linewidth}
            \centering
            \includegraphics[width=\linewidth]{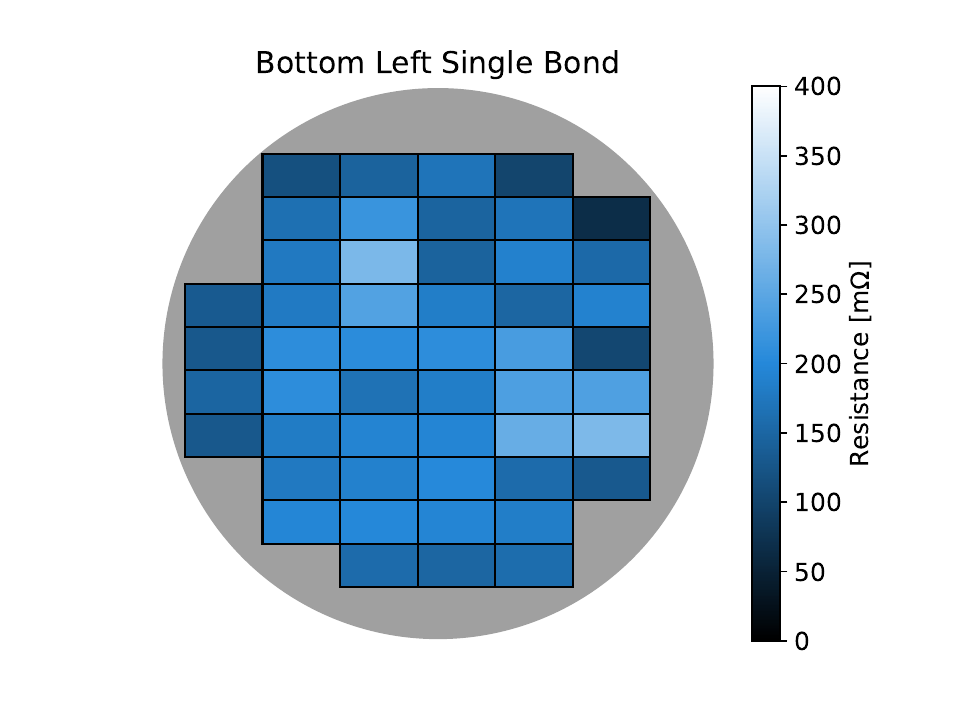}
            \caption{}
        \end{subfigure}
    \end{minipage}
    \hfill
    \begin{minipage}{0.3\textwidth}
        \begin{subfigure}[]{\linewidth}
            \centering
            \includegraphics[width=\linewidth]{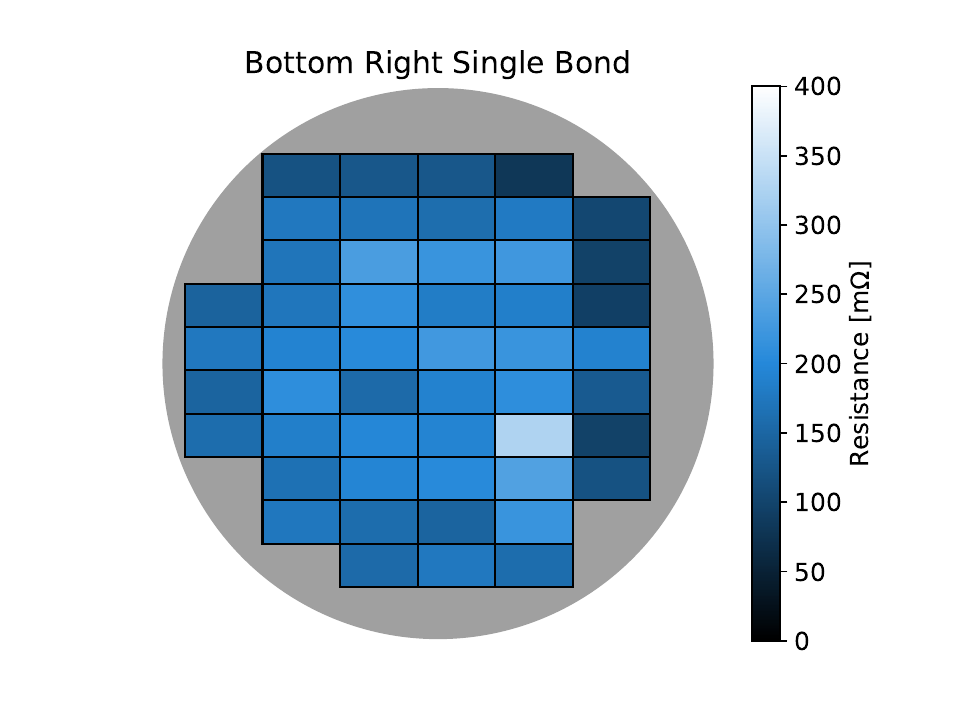}
            \caption{}
        \end{subfigure}
        \bigskip
        \begin{subfigure}[]{\linewidth}
            \centering
            \includegraphics[width=\linewidth]{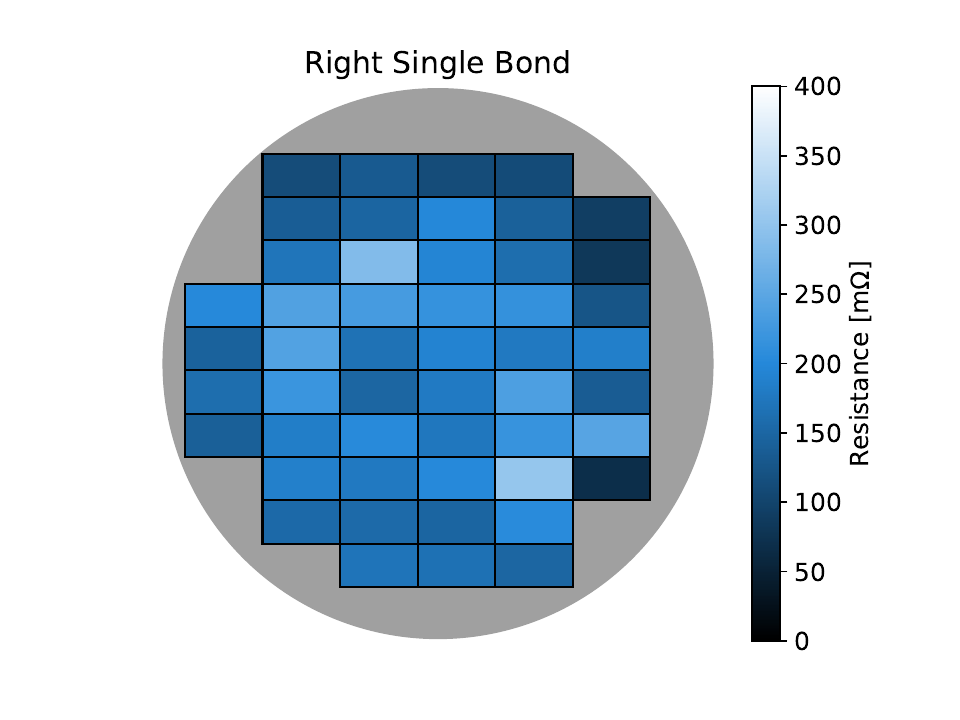}
                    \caption{}
        \end{subfigure}
    \end{minipage}
    \caption{Wafer maps showing the resistances of the five single bond test structures per die on a common color scale from \qtyrange{0}{400}{\milli\ohm}. The resistance of the die in red in the center of the top left structure \autoref{subfig:SBSTopLeftCommScale} could not be measured with the chosen resistance range and is considered a broken bond.}
    \label{fig:SBSwafer_plots}
\end{figure}

\begin{figure}
    \centering
    \includegraphics[width=0.8\linewidth]{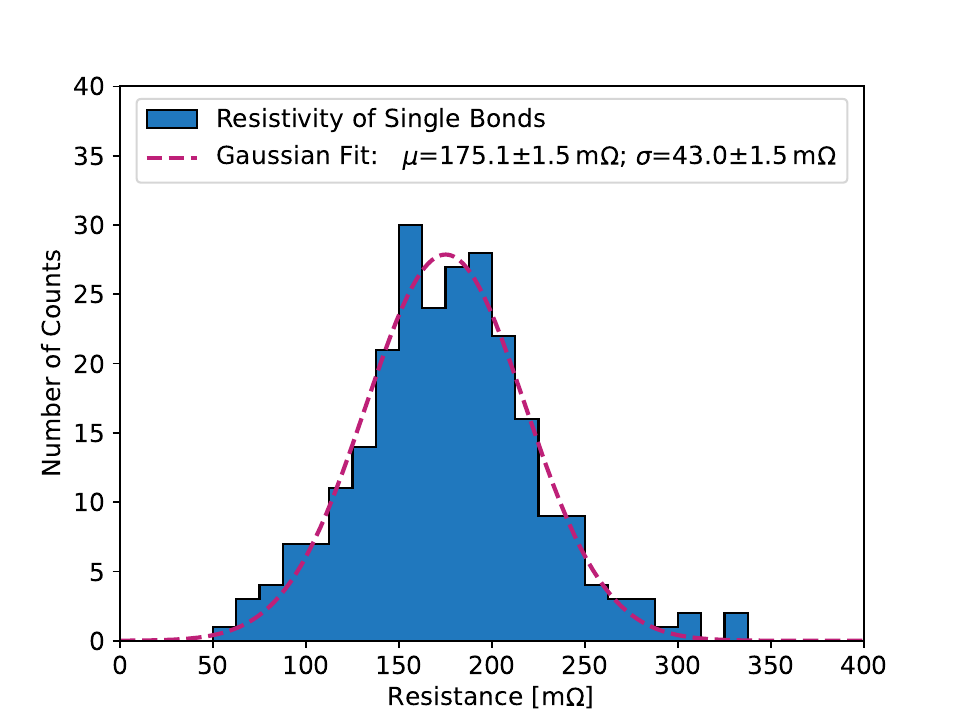}
    \caption{Distribution of single bond resistance values across all five structures. The mean resistance per bond was determined by fitting a Gaussian function to the distribution (dashed line).}
    \label{fig:SBShist_combined}
\end{figure}

Dedicated structures on the DCW stack allow for a resistance measurement between the two wafers through a single interconnect. 
Five single bond structures can be used for measurements on each chip.
They are designed for four-terminal sensing and thus provide four different pads.
The resulting resistances for each of the five structures are shown in \autoref{fig:SBSwafer_plots}. 
Each plot shows the fifty dies of the DCW stack as colored rectangles at their respective relative position on the wafer stack. 
The color indicates the measured mean resistance value in a range from \SI{0}{\milli\ohm} (dark blue) to \SI{400}{\milli\ohm} (light blue). 
Only one of the single bonds was found to be disconnected (red).
\autoref{fig:SBShist_combined} shows the resistance distribution of all single bond structures combined. 
The relative uncertainties of the individual resistance values are up to 3\% for the smallest measured values. 
The mean resistance per bond was determined to be \SI{175.1\pm 1.5}{\milli\ohm} with a standard deviation of \SI{43.0\pm 1.5}{\milli\ohm}.
\subsubsection{Daisy Chains}
\label{subsubsect:DC}
\begin{figure}
    \centering    \includegraphics[width=0.8\linewidth]{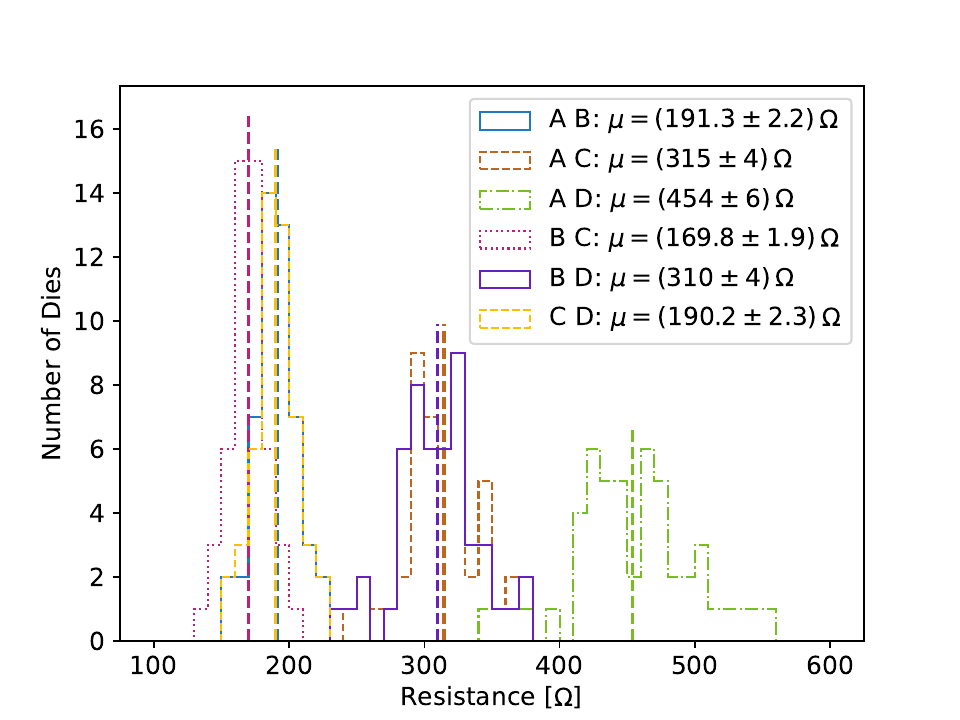}
    \caption{Resistances of all dies for different sections of the center daisy chains. For longer chains the resistance and width of the distribution increases.}
    \label{fig:CenterDCS_hist_combined}
\end{figure}

\begin{figure}[b]
\begin{subfigure}[c]{0.3\textwidth}
    \centering
    \includegraphics[width=\linewidth]{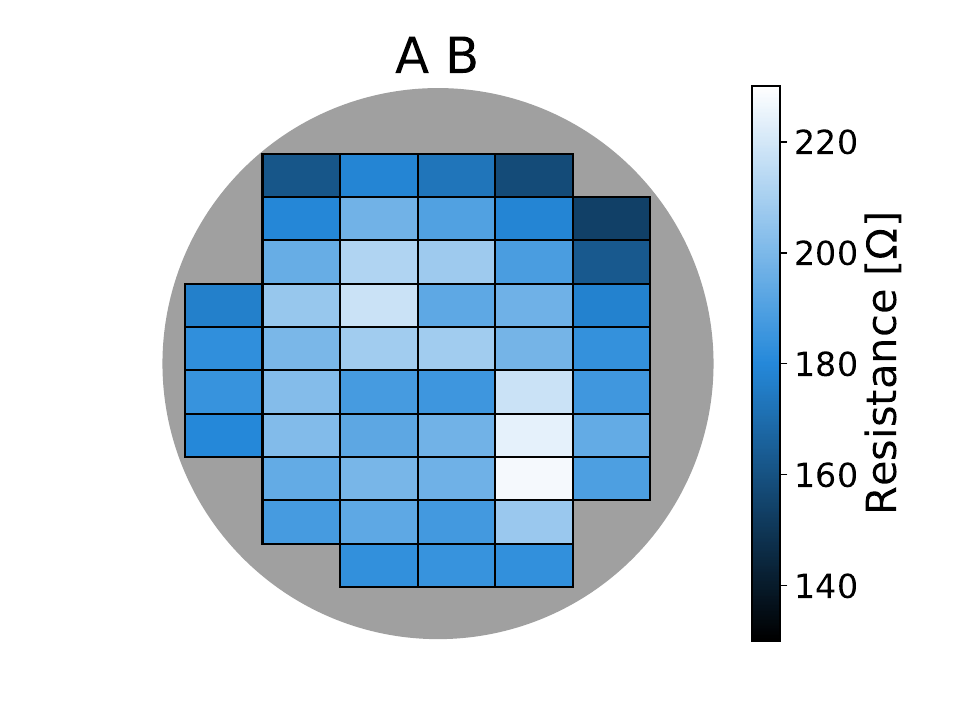}
    \caption{ }
\end{subfigure}
\begin{subfigure}[c]{0.3\textwidth}
    \centering
    \includegraphics[width=\linewidth]{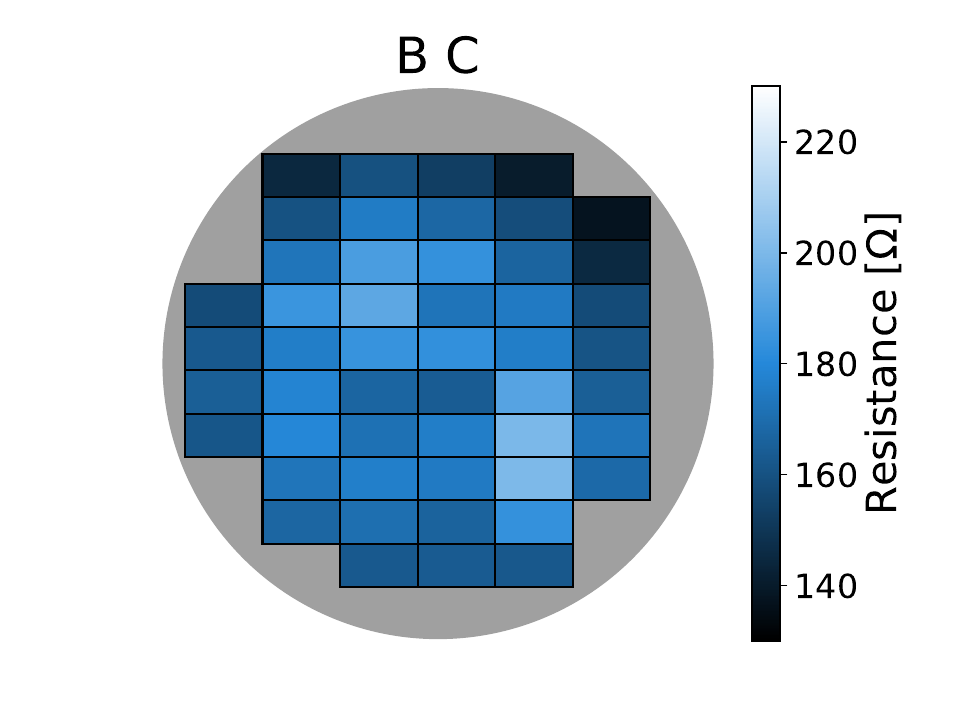}
    \caption{}
\end{subfigure}
\begin{subfigure}[c]{0.3\textwidth}
    \centering
    \includegraphics[width=\linewidth]{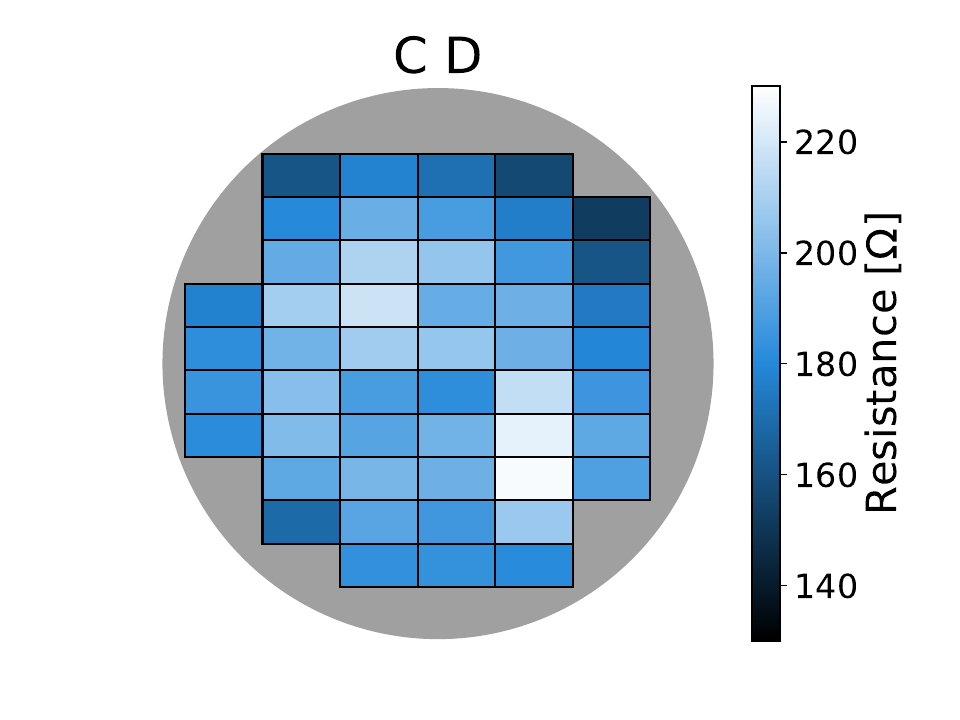}
    \caption{ }
\end{subfigure}
\caption{Measured resistances for the three short sections of the center daisy chain. The color indicates the measured resistance of the respective path on each die. While the absolute resistances differ from die to die the distributions across the wafer are very similar for the different structures.
}
    \label{fig:CenterDCSwafer_plots}
\end{figure}

\begin{figure}
    \centering    \includegraphics[width=0.8\linewidth]{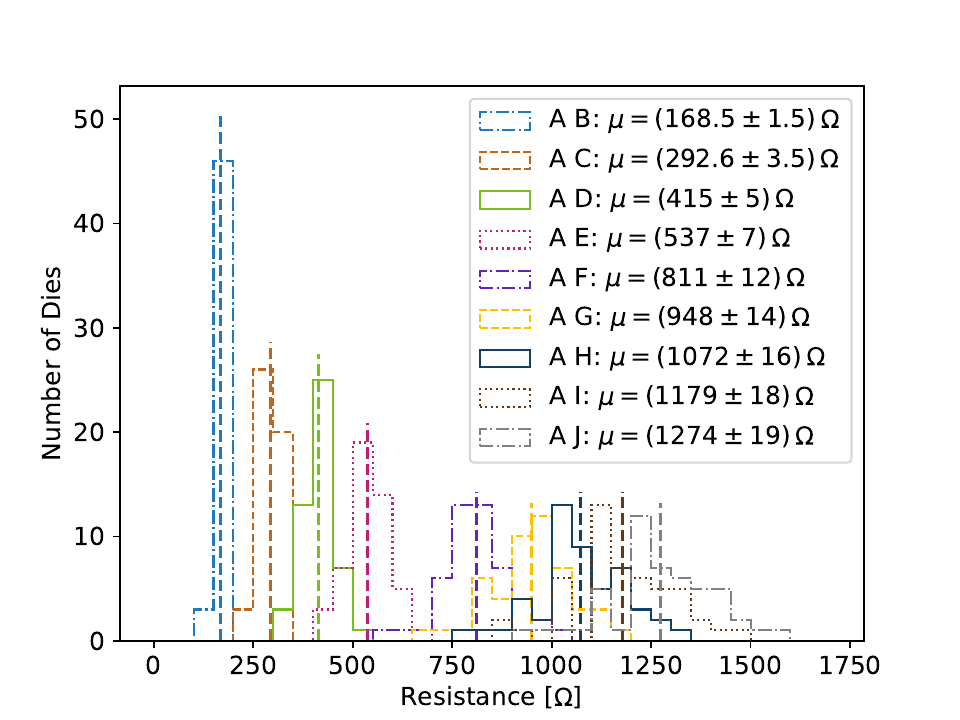}
    \caption{Resistances of all dies for top daisy chain sections of increasing length. The longer the daisy chain the higher the measured resistance and the width of the resistance distribution.}
    \label{fig:TopDCS_hist_combined}
\end{figure}

\begin{figure}
\begin{subfigure}[c]{0.49\textwidth}
    \centering
    \includegraphics[width=0.8\linewidth]{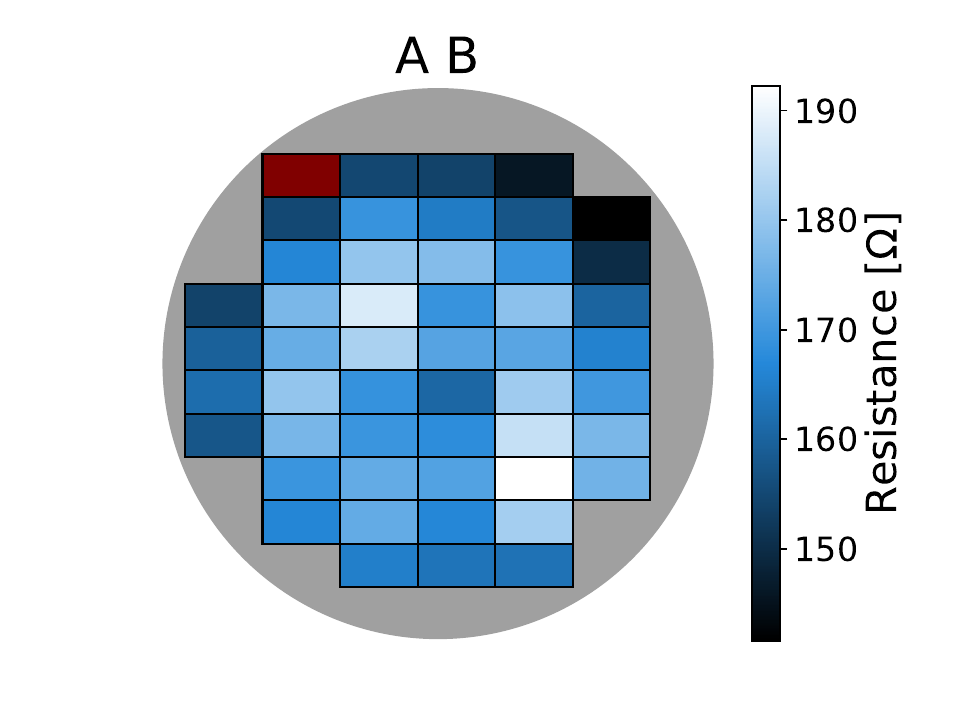}
    \caption{}
    \label{subfig:TopDCSwafer_plots_AB}
\end{subfigure}
\begin{subfigure}[c]{0.49\textwidth}
    \centering
    \includegraphics[width=0.8\linewidth]{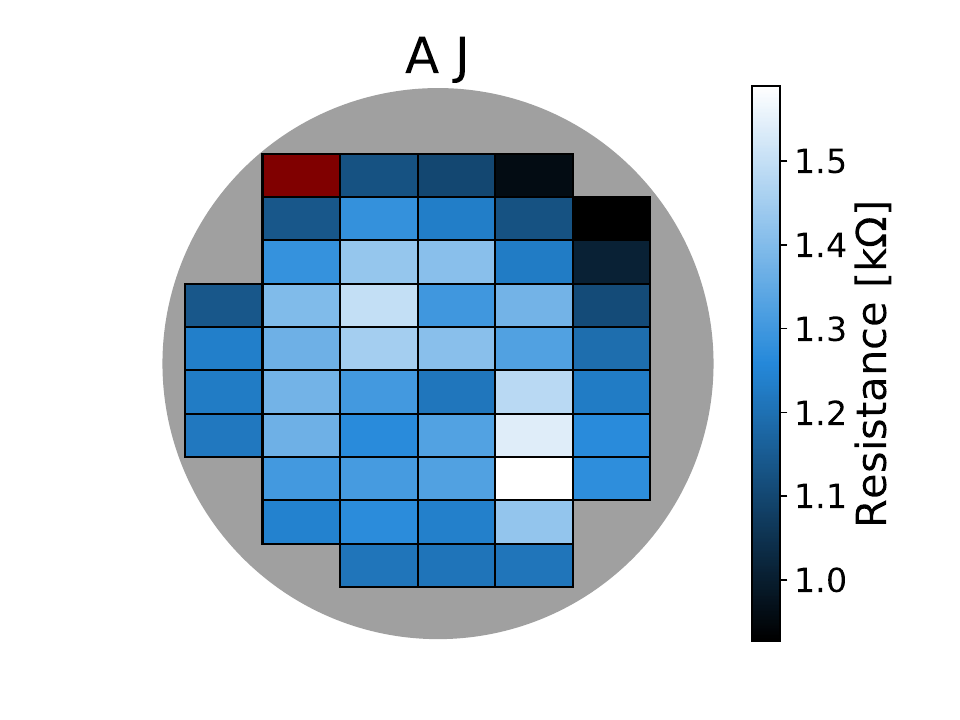}
    \caption{}
    \label{subfig:TopDCSwafer_plots_AJ}
\end{subfigure}
    \caption{Wafer maps showing the measured resistances for a short path between pads A and B (\autoref{subfig:TopDCSwafer_plots_AB}) and the longest path between pads A and J (\autoref{subfig:TopDCSwafer_plots_AJ}). 
    The dies that exceed the measurement range are shown in red and classified as broken daisy chains. 
    Each wafer map is shown in their own resistance range to highlight relative local differences. 
    The structures present in the short daisy chain are similar to the ones of the long daisy chain.}
    \label{fig:TopDCSwafer_plots}
\end{figure}

Each die on the DCW stack contains three different types of daisy chain structures: four corner daisy chains, one center daisy chain and one top daisy chain.
\begin{figure}
    \centering
    \begin{subfigure}[]{0.40\textwidth}
        \centering
        \includegraphics[width=\textwidth]{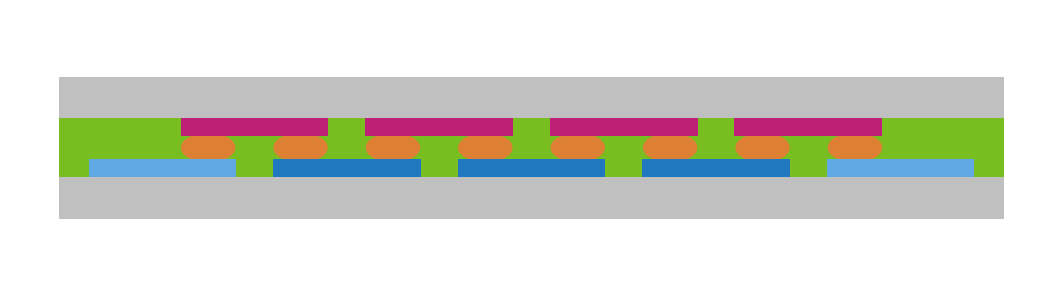}
        \caption{}
        \label{subfig:DC_schematics1D}
    \end{subfigure}
    \begin{subfigure}[]{0.38\textwidth}
        \centering
        \includegraphics[width=\textwidth, angle=180]{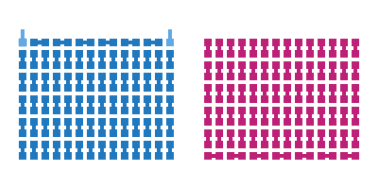}
        \caption{}
        \label{subfig:DC_schematics2D}
    \end{subfigure}
    \begin{subfigure}[]{0.2\textwidth}
        \centering
        \includegraphics[width=\textwidth]{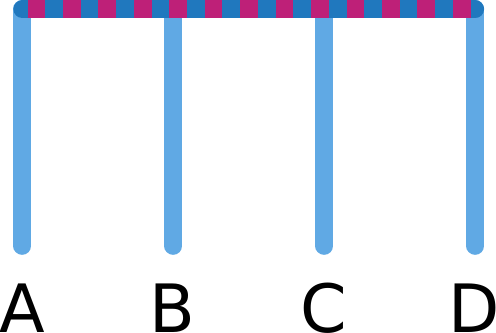}
        \caption{}
        \label{fig:CenterDCSchematic}
    \end{subfigure}
    \caption{Schematics of a daisy-chain structures. 
    \autoref{subfig:DC_schematics1D} shows a schematic of the cross section of a daisy chain on the two bonded wafers (gray) and their respective traces (magenta and blue). 
    The traces are electrically connected via the bonds (orange) and stabilized with a bonding polymer layer (green) and build the DCW stack. 
    The first and last traces lead to the pads (light blue). 
    \autoref{subfig:DC_schematics2D} shows a top view of a daisy chain realized where the pads on one wafer (blue) form a chain when flipped onto the structure of the second wafer (magenta). 
    This structure was realized for the corner daisy chains. \autoref{fig:CenterDCSchematic} shows a sketch of the center daisy chain, which can be contacted at the end of four paths A, B, C and D (light blue) to measure different sections of the daisy chain (magenta, dark blue dashed line)}
    \label{fig:DC_schematics}
\end{figure}
A schematic of a daisy chain is shown in \autoref{subfig:DC_schematics1D}. 
The upper and lower wafer (gray) house traces (blue and magenta) that connect neighboring bonds (orange). 
\autoref{subfig:DC_schematics2D} shows the two matching structures of a two dimensional daisy chain as implemented for the corner daisy chains.
Pads on each end (light blue) can be used for contacting the daisy chains.
A four-point measurement was performed, where two probe needles were placed on each pad for the power and sense lines, respectively. 
This compensates for the resistance of the cables from the multimeter to the pads but not the resistance of the traces from the pads to the daisy chain on the die.
\\
\\
\noindent
The resistances of the two corner daisy-chain structures at the top of the dies are about \SI{75}{\ohm} higher than for the two structures in the bottom part of the dies due to the longer traces from the pads to the structures.
Daisy chains with measured resistances in excess of \SI{500}{\ohm} are considered open.
\\
\\
\noindent
During first measurements, the corner daisy chains on this DCW stack were systematically damaged and a reliable yield can not be calculated.
Independent measurements of a second DCW stack resulted in a measured yield of 99\% with two open daisy-chain structures towards the edge of the DCW stack.
\\
\\
\noindent
The daisy chain in the center of the wafer stack is split into three parts and can be contacted at four different points A, B, C and D along its chain, as shown in \autoref{fig:CenterDCSchematic}.
\autoref{fig:CenterDCS_hist_combined} shows the distributions of the resistances of the different sections and their mean values as dashed vertical lines of the same color. 
Since the chain from pad A to D (light green) is the longest its mean resistance of \SI{454 \pm 6}{\ohm} is the highest. 
The chains from pad A to C (orange) and pad B to D (purple) have the same length and result in similar mean measured resistances of \SI{315 \pm 4}{\ohm} and \SI{310 \pm 4}{\ohm}, respectively. 
The shortest chains from pad A to B (blue), B to C (magenta) and C to D (yellow), again have similar resistance values. 
\\
\\
\noindent
\autoref{fig:CenterDCSwafer_plots} shows the wafer maps of the resistances of the three short chain sections on a common color scale. 
The pattern in resistance of neighboring dies changes smoothly over the DCW stack. Two local maxima in the upper left and lower right of the DCW stack are present, while dies close to the upper right edge have lower resistances. 
This structure is very similar in all three daisy chain sections. 
However, since this structure can not be observed in the single bond structures shown in \autoref{fig:SBSwafer_plots} it likely does not stem from the interconnections but from the circuit traces.
The resistances on all dies is below \SI{600}{\ohm}, which indicates no broken bonds along any of the center daisy chains. The second DCW stack showed one disconnected center daisy chain.
\\
\\
\noindent
The daisy chain in the top part of the dies is split into nine sections of different lengths with contactable pads labeled A to J. 
\autoref{fig:TopDCS_hist_combined} shows the resistance distributions across all dies for different path lengths, always starting with the first pad, and their respective mean values as dashed vertical lines of the same color.
The systematic increase of the resistance with the daisy chain length can be seen. However, the individual sections of the daisy chains have different lengths, resulting in non-linear increase of the resistance, noticeable especially by the gap between the measurement from A to E and A to F.
For longer daisy chains the distribution broadens. 
\\
\autoref{fig:TopDCSwafer_plots} shows the results of two exemplary resistance measurements. 
\autoref{subfig:TopDCSwafer_plots_AB} shows a short daisy chain path from pads A to B and  \autoref{subfig:TopDCSwafer_plots_AJ} shows the complete daisy chain from pads A to J. 
It shows a similar distribution of resistance values across the wafer stack for both measurements, which is also similar to the distribution of center daisy chain resistances shown in \autoref{fig:CenterDCSwafer_plots}. 
For one die in the upper left, the resistance exceeds the chosen measurement range for measurements from pad A to all other pads, resulting in it being considered disconnected. For the second DCW stack two dies show a disconnection.

%% file: SensorWafer.tex
\subsection{Design of the sensor wafer}
\label{subsect:SW_properties}
The first prototype of wafer-to-wafer bonded ultra-thin hybrid pixel detectors will be based on large-area passive CMOS sensors designed specifically for this project~\cite{PhDYannick}.
   \begin{figure}
       \centering
       \includegraphics[width=0.8\linewidth]{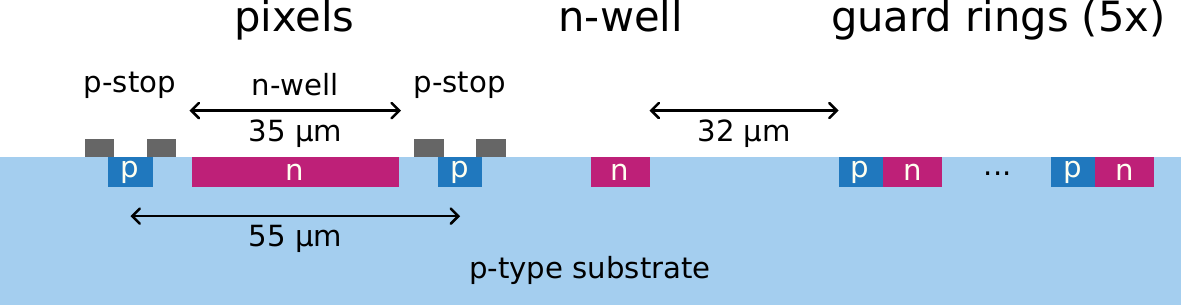}
       \caption{Schematic of the edge region of a sensor chip with the outer most pixel, the n-well and the five guard rings.}
       \label{fig:SensorSchematics}
   \end{figure}
     A schematic of the sensors edge is shown in \autoref{fig:SensorSchematics}. 
     The sensor is realized as an n-in-p pixel matrix with a pixel pitch of \SI{55}{\micro \meter} by \SI{55}{\micro \meter}. 
     Each pixel consists of a \SI{35}{\micro \meter} wide charge collecting n-well isolated by a p-stop implant. 
     The sensitive area of a sensor is confined by an n-well around the pixel matrix. 
     Five n-p guard rings around the n-ring are needed for a smooth voltage drop between the sensitive area and the edge of the die to improve the breakdown behavior. 
     For testing purposes before the bonding of the sensor a bias grid, including a bias resistor, has been realized. Each die provides two sets of probe pads for the bias grid, the n-ring and guard rings, respectively.\\
    \\
    \noindent
    Timepix3 read-out wafers will be used as front-end of the sensor~\cite{Timepix3}.
    The sensor wafer needs to fit the size and matrix of a single read-out chip as well as the whole Timepix3 wafer layout.
    Therefore, the sensor wafer was designed as a 200mm wafer with 105 usable dies of the same size and positioning on the wafer as for the Timepix3 read-out chip wafers.
    Each die has a size of \SI{14.2}{\milli \meter} by \SI{16.69}{\milli \meter}.
    While the Timepix3 chip features 256x256 pixels per die with a pitch of \SI{55}{\micro \meter}, the sensors only provide 256x246 pixels due to space constraints in x-direction because of the need for n- and guard rings.
    The lower part of the die is unstructured, where the end-of-column logic and the wire bond pads are located on the Timepix3. 
    For traditional bump bonding the sensor die is usually smaller than the read-out chip, leaving the wire bond pads accessible. Due to the wafer-to-wafer bonding, the wire bond pads are covered by the sensor wafer. 
    Through-silicon vias will be used to route the signals to the backside of the Timepix3 chips.
    \\
    \\
    \noindent
    The sensor wafer was produced using an LFoundry \SI{150}{\nano\meter} process. 
    After pixel implantation on the front side, the wafer was thinned from the back to a thickness of \SI{150}{\micro \meter}.
    A backside metallization was applied for contacting the bias voltage via the chuck of the wafer prober.
    
\subsection{Experimental Investigations of the Sensor Wafer Quality}
    \label{subsect:SensorQuali}
    \noindent
    Just like for the DCW stack, both sensor wafers were measured using the semi-automatic wafer prober MPI TS3500-SE.
    The IV-curves of the prototype sensors were measured using a Keithley 2410 SourceMeter.
    The bias voltage was applied via the chuck of the wafer prober to the backside of the wafer.   
    A hardware current limit of $I_{\text{hw}}=\SI{2}{\micro \ampere}$ was set to protect the sensor for the measurement of the IV-curves.
    Additionally, an HP4284A LCR-meter was used for the CV-curves.

    \subsubsection{IV-Characteristics}
    \label{subsubsect:IV}
\begin{figure}
        \centering
        \includegraphics[width=0.6\linewidth]{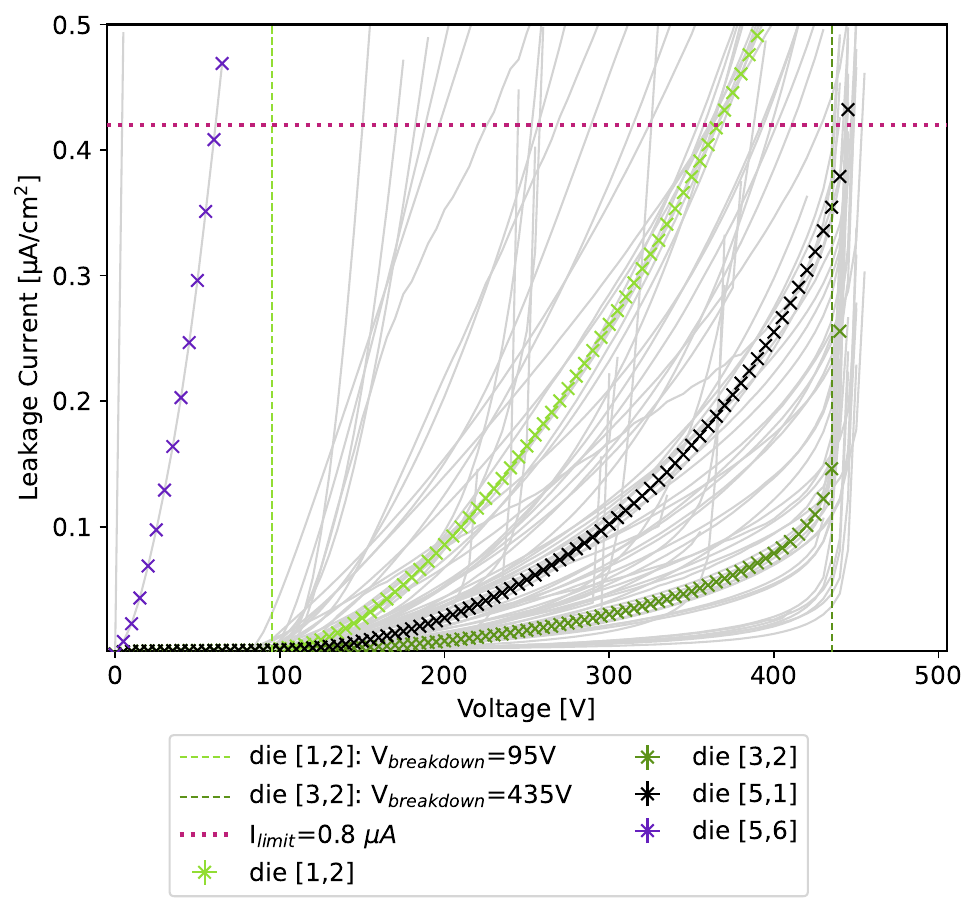}
        \caption{Measured leakage current in dependence of the applied bias voltage for sensor wafer 1. 
        Four example IV-curves shown as data points. 
        The data set shown in violet exceeds the soft current limit (magenta dashed line) before reaching a bias voltage of \SI{150}{V}. 
        For the IV-curves for die [1,2] (light-green) and [3,2] (dark-green) the breakdown voltages have been determined, shown as dashed vertical lines. 
        For die [5,1] (black) the determination of the breakdown voltage was not possible.}
        \label{fig:IV_Curve_Example}
    \end{figure}

\begin{figure}
        \centering
        \begin{subfigure}{0.49\linewidth}
            \centering
            \includegraphics[width=\linewidth]{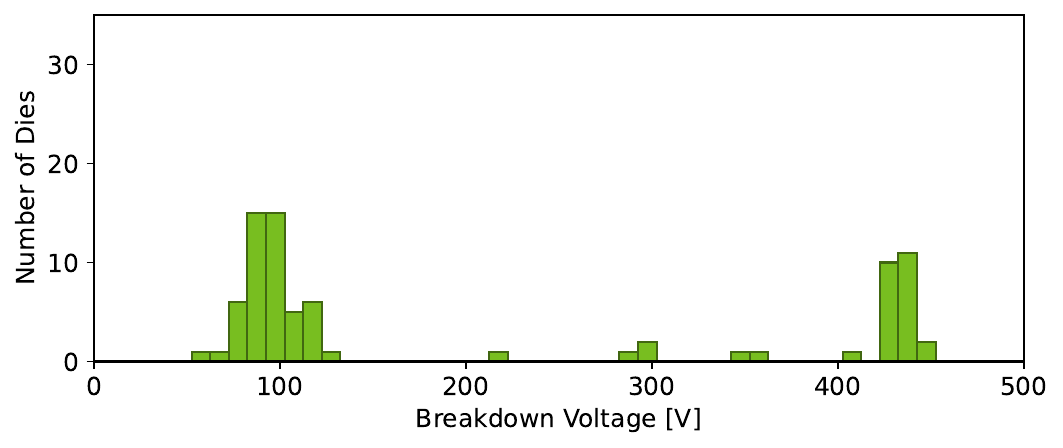}
            \caption{sensor wafer 1}
            \label{subfig:IV_Breakdownhist_SW1}
        \end{subfigure}
        \begin{subfigure}{0.49\linewidth}
            \centering
            \includegraphics[width=\linewidth]{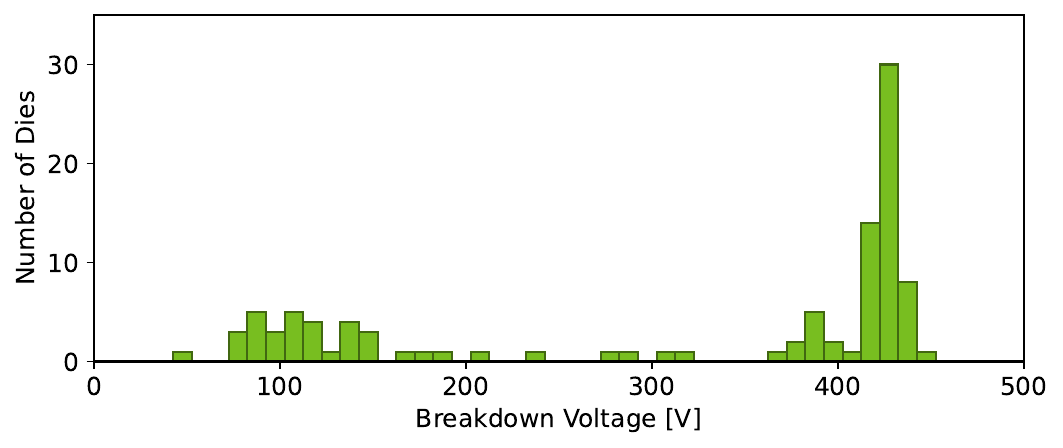}
            \caption{sensor wafer 2}
            \label{subfig:IV_Breakdownhist_SW2}
        \end{subfigure}
        \caption{Histograms of the breakdown voltages for sensor wafer 1 (\autoref{subfig:IV_Breakdownhist_SW1}) and sensor wafer 2 (\autoref{subfig:IV_Breakdownhist_SW2}). 
        Both distributions show two distinct groups of dies with low and high breakdown voltages.}
        \label{fig:IV_Hists}
    \end{figure}

    \autoref{fig:IV_Curve_Example} shows all IV-curves of sensor wafer 1 as gray lines, as well as measurement points for four exemplary curves. 
    A sensor die is classified as good if its leakage current does not exceed the soft current limit of $I_{\text{limit}}=\SI{0.8}{\micro \ampere}$ up to an applied bias voltage of \SI{150}{\volt}.
    The violet points in \autoref{fig:IV_Curve_Example} highlight a die which exceeds the soft current limit, shown as a magenta dotted line, below \SI{150}{\volt}. \\
    \num{85} out of \num{105} dies on sensor wafer 1 and \num{97} of \num{102} dies on sensor wafer 2 are functional.\\
    Sensor breakdown is defined as an increase in leakage current in excess of 20\% for a \SI{5}{\volt} increase of bias voltage.
    The light green measurement points show the IV-curve of a die, whose breakdown voltage is below \SI{100}{\volt}. The breakdown voltage is marked as a dashed vertical line of same color.
    The breakdown voltage of the dark green measurement points is \SI{435}{\volt}. 
    For a measurement with a curvature in between the two green curves, shown in black, the criterion for a breakdown voltage is never satisfied up to the hardware limit and a breakdown voltage can not be determined. 
    This was the case for \num{5} dies on sensor wafer 1 and none on sensor wafer 2.
    \\
    \\
    \noindent
    The resulting breakdown voltage distributions are shown in \autoref{subfig:IV_Breakdownhist_SW1} for sensor wafer 1 with a total of \num{80} dies and in \autoref{subfig:IV_Breakdownhist_SW2} for sensor wafer 2 and \num{102} dies, respectively.
    Both wafers show two distinct groups of dies for which the breakdown voltages lie either around \SI{100}{\volt} or above \SI{400}{\volt}. 
    The low breakdown voltage is most likely caused by a too shallow backside implantation which leads to a diffusion of the metallisation beyond the backside implantation into the p-doped substrate. As soon as the depletion zone reaches this region a current is flowing through the backside, showing in an early breakdown close to the depletion voltage (see. \ref{subsubsect:CV}) \cite{PhDYannick}.
    The breakdown voltage shows no systematic pattern across the wafer.
    
\subsubsection{CV-Characteristics}
    \label{subsubsect:CV}
\begin{figure}
        \centering
        \includegraphics[width=0.5\linewidth]{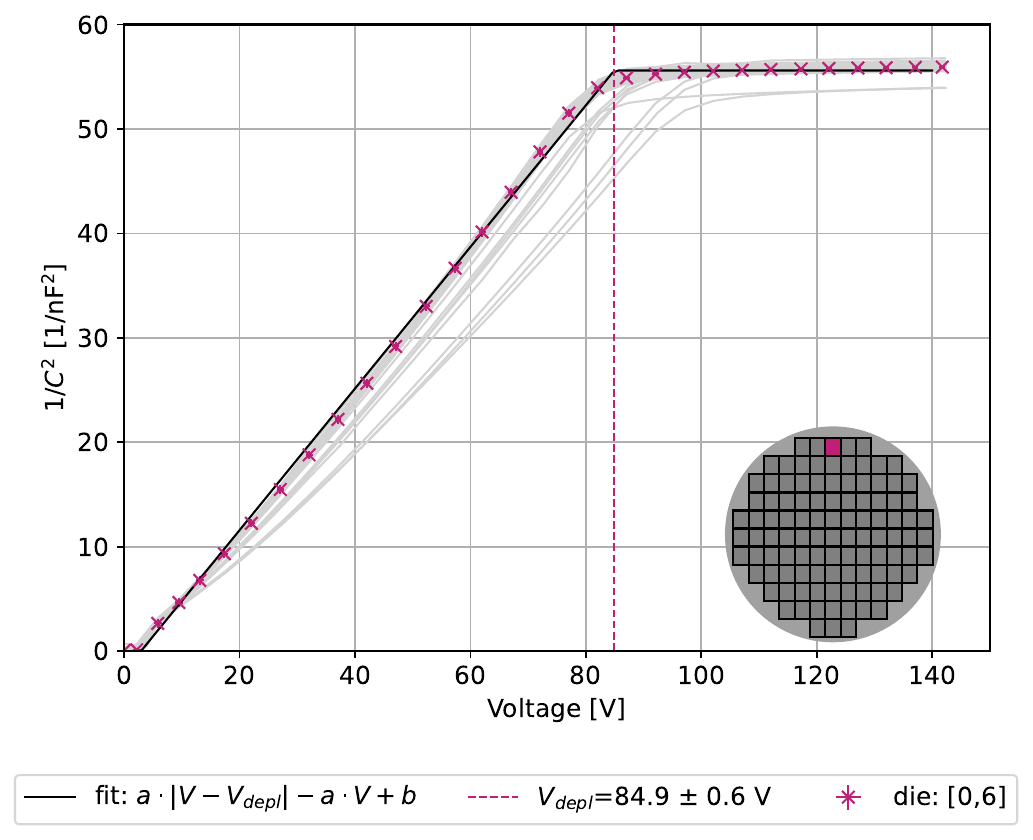}
        \caption{Measured capacitance $C$ in dependence of the applied bias voltage displayed as $\frac{1}{C^2}$. 
        The curves of all measured dies are shown in light gray. 
        An example measurement for die [0,6] is highlighted in magenta with its fit in black and the determined depletion voltage as a vertical magenta dashed line.}
        \label{fig:ExampleCVSW1}
    \end{figure}

\begin{figure}
        \centering
        \includegraphics[width=0.5\linewidth]{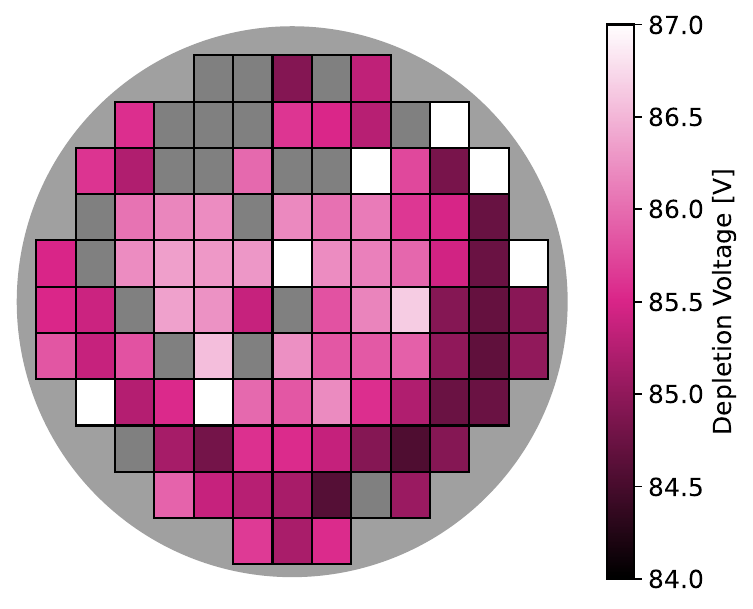}
        \caption{Wafer map of the depletion voltages in the range \qtyrange{84}{87}{\volt}. 
        Dies that were not measured are shown in gray. 
        Dies exceeding this range are shown in yellow. 
        A tendency of higher voltages towards the center of the wafer can be observed. }
        \label{fig:WaferplotCVSW1}
    \end{figure}

    For sensor wafer 1 the capacitance was measured on the \num{85} dies that did not exceed the current limit up to a bias voltage of \SI{150}{\volt}. 
    \autoref{fig:ExampleCVSW1} shows the inverse squared of the capacitance for all measured dies in gray and and example measurement as magenta data points. 
    For low bias voltages, the inverse squared capacitance increases linearly with the applied voltage up to the depletion voltage $V_{\text{depl}}$, where the sensor is fully depleted.
    The depletion voltage was determined by a fit with the free parameters $a$ and $b$:
    \begin{equation}
        C^{-2} = a \cdot | V - V_{\text{depl}} | - a \cdot V + b
    \end{equation}
    In \autoref{fig:ExampleCVSW1} the fit is shown as a black line and the determined depletion voltage as a dashed vertical line. 
    \begin{figure}
        \centering
        \includegraphics[width=0.5\linewidth]{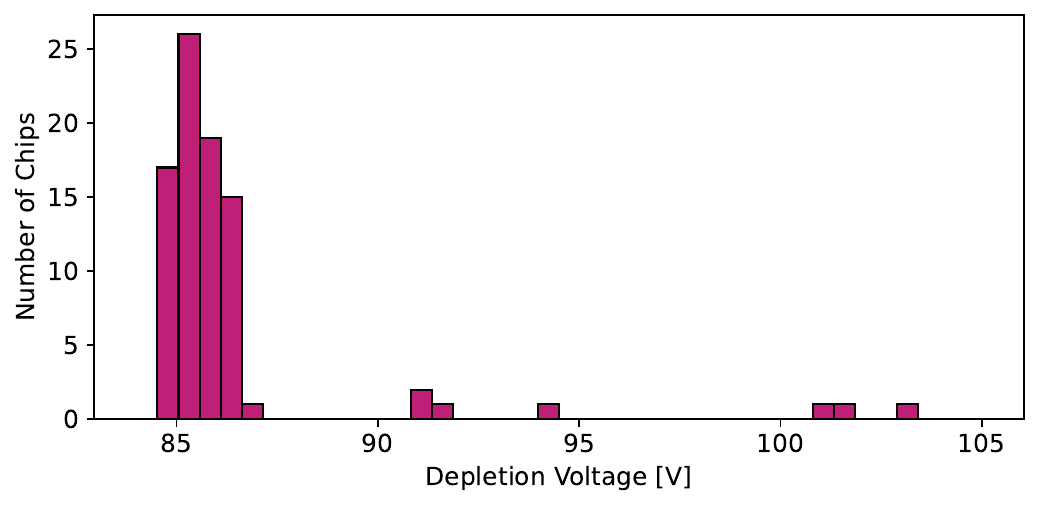}
        \caption{Histogram of the depletion voltages for sensor wafer 1.}
        \label{fig:HistCVSW1}
    \end{figure}
    The measured depletion voltages are shown in \autoref{fig:HistCVSW1}.  
    Most dies show a depletion voltage between \SI{84.5}{\volt} and \SI{87.0}{\volt}. 
    \autoref{fig:WaferplotCVSW1} shows the distribution of depletion voltages across the wafer. 
    Not measured dies are shown in gray. 
    Dies with a depletion voltage above \SI{87}{\volt} are shown in white. 
    A tendency towards higher depletion voltages is clearly visible in the middle of the wafer.
    For operation as a particle detector the bias voltage is chosen to be at least the depletion voltage to utilize the whole depleted thickness of the sensor as sensitive volume. However, this is only feasible if the depletion voltage is smaller than the breakdown voltage to not damage the sensor during operation. For \num{13} dies $V_{\text{depl}}<V_{\text{breakdown}}$, which reduces the number of utilizable dies to \num{72}.

%% file: Conclusion.tex
This paper shows the current progress towards ultra-thin wafer-to-wafer bonded hybrid pixel detectors. 
Two daisy-chain wafer stacks were investigated independently to determine the bump bond quality in terms of yield and single bump resistance.
We have shown yields of \SI{99.2}{\%} and \SI{99.6}{\%} for the single bond structures, \SI{100}{\%} for the center daisy chain, and \SI{96}{\%} and \SI{98}{\%} for the top daisy chain.
The corner daisy chains on one of the wafer stacks were damaged during measurement, while the second wafer stack showed a yield of \SI{96}{\%}.
The measurements show that possible bonding defects tend to be localized near the edge of the wafer.
\\
\\
\noindent
Sensor wafers that match the Timepix3 read-out chip wafers were designed, produced and investigated in preparation for hybridization. 
In terms of breakdown voltage, two classes of sensors can be identified.
For one, a sharp increase of the leakage current can be observed for bias voltages of about \SI{100}{\volt}, slightly above the depletion voltage.
This is likely a result of parasitic current flowing  through the backside implant when the depletion region reaches the backside of the sensor.
For the majority of dies, however, breakdown voltages in excess of \SI{430}{\volt} were measured.
Sensor wafer 1 was tested further and shows an average full depletion voltage of \SI{86.43\pm0.07}{\volt}.
Defining a utilizable sensor as described in section~\ref{subsubsect:CV}, sensor wafer 1 shows a yield of \SI{69}{\%}.
\\
\newline
Our work demonstrates that the designed sensor wafers work as intended and are utilizable for wafer-to-wafer bonding. The developed polymer-metal-hybrid bonding process shows an excellent yield and can be used to reliably interconnect wafers.
\\
\newline In the next step, the dedicated sensor wafers will be bonded to Timepix3 read-out chip wafers using this technique and thinned down. Then, we will test 
the ultra-thin, low-mass hybrid pixel detectors for their use in high-energy physics experiments.